\documentclass[12pt, final]{iopart}

\usepackage{etoolbox}
\usepackage[utf8]{inputenc}

\usepackage{iopams}

\expandafter\let\csname equation*\endcsname\relax

\expandafter\let\csname endequation*\endcsname\relax

\usepackage{mathrsfs,mathtools,bbold}
\usepackage{amsmath,amssymb}
\usepackage{amsfonts}
\usepackage{float, cite}
\usepackage[colorlinks,linktocpage]{hyperref}
\usepackage[dvipsnames]{xcolor}
\usepackage{physics}
\usepackage{slashed}
\usepackage[singlelinecheck=false 
]{caption}
\usepackage[flushleft]{threeparttable}

\hypersetup{colorlinks=true,linkcolor=blue,citecolor=blue,urlcolor=blue}

\begin{document}
\title{Search for electric dipole moment}
\author{V. G. Baryshevsky$^1$ and P. I. Porshnev$^2$}
\address{$^1$ Institute for Nuclear Problems, Belarusian State University, Minsk, Belarus}
\address{$^2$ Past affiliation: Physics Department, Belarusian State University, Minsk, Belarus}
\vspace{10pt}
\begin{indented}
\item[]October 2020
\end{indented}

\begin{abstract}
The outstanding progress has been made in reducing the upper bounds on EDM of several particles. Even if significant challenges must be overcome to further improve these limits, it is still one of the best chances to detect new type of interactions beyond the standard model. Analyzing several examples, we highlight a common thread that is visible in different set-ups used  for the EDM detection. The electric dipole moment is one of the clear consequences of CP- or T-violating interactions, however it is not the only one. These symmetry-violating interactions enable extra phenomena that unavoidably accompany the EDM-induced spin precession, and they must be taken into account in planning and executing sensitive experiments. After reviewing three typical cases, we suggest conditions for improving the sensitivity of detecting the intrinsic EDM.
\end{abstract}

\vspace{2pc}
%
%
%

\date{\today}

\section{Introduction}
In this brief review, we focus on one specific challenge of electric dipole moment (EDM) experiments. A violation of symmetry that enables the existence of EDM $d$ also allows to introduce new terms into Lagrangian or Hamiltonian, which in turn enter the equations for observables. The key feature of these $CP$- or $T$-odd terms is that they mimic or imitate the influence of EDM itself.  Therefore, the effective EDM $d_{eff}$ that is observed typically includes additional contributions  in the form
\begin{equation}
		W_{edm} = - \vec{d}_{eff}\cdot \vec{E} = - \qty[\vec{d}+ \vec{d}_{ind}(f)]\cdot \vec{E}\,.
\end{equation} 
where the intrinsic dipole moment $\vec{d}=d \vec{S} $ is aligned along the particle spin $\vec{S}$, $\vec{E}$ is an electric field,   and $W_{edm}$ is the energy of dipole interaction with field $\vec{E}$.  While the EDM coefficient $d$ itself is the fundamental constant, the additional contribution $\vec{d}_{ind}(f)$ is of dynamic nature, since it depends on external factors $f$. It allows to clearly distinguish two contributions from each other, in principle. 

To illustrate this interplay, consider an atom under influence of laser field \cite{baryshevsky_parity_1993, baryshevsky_phenomenon_1999}. The Schroedinger equation that describe this problem can be given as 
\begin{equation}
	i\hbar \pdv{\psi}{t} = H\psi=\qty(H_0 - \vec{d}\cdot \vec{E} - \vec{\mu}\cdot \vec{B} + U)\psi\,,
\end{equation}
where $H_0$ is the atomic Hamiltonian in the absence of external field,  $\vec{B}$ is the magnetic field, $d$ and $\mu$ are the atom electric and magnetic moments respectively. The additional energy $U$ of atom-field interaction is originated by the atom polarizability 
\begin{equation}\label{eq9}
	U = U_{sym}+U_{mix}=- \frac{1}{2} \qty(  \hat \alpha_{ik}E_iE_k^* + \hat \beta_{ik }B_iB_k^*) - \frac{1}{2} \qty(\hat \alpha^{\prime}_{ik}B_iE_k^* +\hat \beta^\prime_{ik}E_iB_k^*) \,.
\end{equation}
where we neglect the non-Hermitian parts of polarizability operators, see details in \cite{baryshevsky_parity_1993}. The tensor operators $\hat \alpha_{ik}$ and $\hat\beta_{ik}$ describe the electric and magnetic polarizabilities respectively.

The primed polarizabilities are T-odd  and P-odd  \cite{baryshevsky_parity_1993, baryshevsky_phenomenon_1999, baryshevsky_time-reversal-violating_2000, baryshevsky_phenomenon_1999-1}, and they mix the actions of electric and magnetic fields. The polarizabilities are originated by symmetry-violating interactions of atomic electrons with nucleus nucleons which might be dependent on nuclear spin. The primed polarizabilities could exist even if electrons and nucleus do not possess EDMs.  They are related to each other by Hermitian conjugation, $(\alpha^{\prime}_{ik})^\dagger = \beta^\prime_{ki}$, see the definitions in \cite{baryshevsky_parity_1993, baryshevsky_phenomenon_1999, baryshevsky_time-reversal-violating_2000, baryshevsky_phenomenon_1999-1, baryshevsky_phenomenon_2004,baryshevsky_high-energy_2012 }, which should also be expected from the structure of corresponding terms in \eqref{eq9}.   A convenient scaling of their scalar parts can be given as  $\beta^\prime_s {\sim}\alpha\,\alpha_s \eta$ where $\alpha$ is fine structure constant, $\alpha_s$ is the scalar part of regular $T$- and $P$-even electric polarizability, and $\eta$ is the measure of mixing of levels with opposite parity \cite{baryshevsky_time-reversal-violating_2004}.  

The expression \eqref{eq9} and corresponding expressions below \eqref{eq5}-\eqref{eq6} do not include $P$-odd $T$-even polarizabilities which cause a rotation of light polarization plane by neutral  $P$-odd $T$-even currents \cite{khriplovich_parity_1991}. This effect was experimentally confirmed in Novosibirsk. We do not consider them here, since they do not contribute into either electric or magnetic moments.

Ignoring the symmetry-violating polarizabilities, the regular moments  $(d,\mu)$ and unprimed polarizabilities $(\alpha_{ik}, \beta_{ik})$ ensure the clear distinction between actions of $\vec{E}$ and $\vec{B}$
\begin{gather}
\begin{aligned}\label{eq5}
	&\pdv{(H_0+U_{sym})}{E_i} = - d S_i - \frac{1}{2}\hat \alpha_{ik} E_k^* =- (\hat d_{eff})_i\,,\\[1ex]
	&\pdv{(H_0+U_{sym})}{B_i} = - \mu S_i-\frac{1}{2}\hat \beta_{ik} B_k^*=- (\hat \mu_{eff})_i\,, 
\end{aligned}
\end{gather} 
where we treat the fields and their complex-conjugated ones as independent. The result \eqref{eq5} justifies the names for electric and magnetic moments $(d,\mu)$ and polarizabilities $(\alpha_{ik}, \beta_{ik})$ correspondingly.

By definition, two moments distinguish the contributions into energy between $\vec{E}$ and $\vec{B}$. Even if the effective moments $(d_{eff},\mu_{eff})$, which are given by the right-hand sides of \eqref{eq5}, become dependent on their corresponding fields, there is no mixing between two actions.\hspace{-1ex}
	\footnote[7]{We assume that measurements are performed in a fixed frame, hence ignoring that fields can be transformed into each other by boosts. }
Now, if we include the symmetry-violating polarizabilities
\begin{gather}
\begin{aligned}\label{eq6}
	&\pdv{\hat H}{E_i} = - d S_i - \frac{1}{2}\hat\alpha_{ik} E_k^*- \frac{1}{2}\hat\beta^\prime_{ik} B_k^*=- (\hat d'_{eff})_i\,,\\[1ex]
	&\pdv{\hat H}{B_i} = - \mu S_i-\frac{1}{2}\hat\beta_{ik} B_k^*- \frac{1}{2}\hat\alpha^\prime_{ik} E_k^*=- (\hat \mu'_{eff})_i\,, 
\end{aligned}
\end{gather} 
the boundary between actions of $\vec{E}$ and $\vec{B}$ gets blurrier. The primed polarizabilities, which are $P$- and $T$-odd, mix electric and magnetic contributions into the system energy. The scalar parts of these polarizabilities must be pseudoscalar quantities, since  $\vec{E}\cdot\vec{B}$ is the Lorentz pseudoscalar, while the corresponding total contribution into $U$ must remain $P$- and $T$-even.  

We could immediately highlight two general features that are visible across different experimental set-ups used in the EDM experiments. First, the existence of additional terms in the right-hand sides of \eqref{eq6} makes the physics much richer. Even if both EDM constant $d$ and regular electric polarizability are zero, a nonzero $\beta^\prime_{ik}$ will make the system behave in electric field as if it possesses an electric dipole moment!  The interplay between electric and magnetic moments from the one hand and symmetry-violating polarizabilities from the other hand can be stated as follows.  The existence of EDM means that a particle has some distributed electric field which by means of its $P/T$-odd polarizability contributes into its magnetic moment. By reciprocity, the existence of magnetic moment contributes into its  EDM by means of the $P/T$-odd polarizability. It has been shown for composite systems, including atoms, nuclei and baryons, see \cite{baryshevsky_high-energy_2012, baryshevsky_time-reversal-violating_2004}. As we mentioned before, $P$ and $T$-odd polarizabilities blur the distinction between electric and magnetic properties that are described by corresponding moments.  Second, both proper electric and magnetic moments $(d,\mu)$ are fundamental constants, while the additional terms in the right-hand sides of \eqref{eq6} depend on external factors.  In principle, it makes the contributions of intrinsic moments distinguishable from the effects of field-dependent polarizabilities. Changing the relative field orientation, averaging in space or time, or adjusting field frequencies are regular experimental methods to either suppress or magnify contributions from the field-dependent terms.

Several papers \cite{chupp_electric_2015, cesarotti_interpreting_2019} reviewed different aspects of this challenge of EDM experiments before, to cite two recent ones. An interplay between $d_e$ proper, the $P/T$-odd pseudoscalar and tensor electron-nucleon interactions that are characterized by strengths $C_S$ and $C_T$ respectively, and the isoscalar and isovector pion-nucleon interactions were systematically studied in \cite{chupp_electric_2015} as multiple contributions into the experimental atomic, hadronic and nuclear EDMs. A possibility that the electron EDM measured in ThO experiments is not the $d_e$ proper but the $CP$-odd electron-nucleon interaction
\begin{equation}
	-iC_S\, \bar{e} \gamma^5 e\, \bar{N}N
\end{equation}
was studied in \cite{cesarotti_interpreting_2019}. The effective EDM was given as 
\begin{equation}\label{eq15}
	d_{ThO} = d_e + k C_S\,,
\end{equation}
where the coefficient $k$ depends on ratio of atomic and nuclear radii and atomic factor $Z\alpha$, see its definition in \cite{chupp_electric_2015}. The recent study \cite{flambaum_sensitivity_2020} on EDM of paramagnetic atoms derived a complicated expression for the effective EDM which is still structurally similar to \eqref{eq15}. They also remarked that the $d_e$ contribution into atomic EDMs is subdominant to semileptonic operators $(C_S)$ for hadronic origins of $CP$ violation.  As a side note for the future discussion, we remark that the electron pseudoscalar $i\bar{e} \gamma^5 e$ is one of the origins of this additional contribution into the effective EDM, while $C_S$ characterizes the strength of the corresponding $CP$-odd interaction.

Matrix elements in equations for observables will now  include mixed terms which relative strength must be carefully analyzed and taken into account. One of the striking effects of such a mixing  is the generation of electric field by applying a magnetic one \cite{baryshevsky_phenomenon_2004, baryshevsky_high-energy_2012};  vice versa, a magnetic field will be generated upon applying external electric one under proper conditions. This case is reviewed next.   We discuss two more cases after it  and suggest ways to improve the sensitivity of future experiments.

\section{Field generation}
Consider first an approach to detect the EDM in solid state samples. The polarizability terms that mix contributions from electric and magnetic fields can actually enable the EDM detection by following mechanism \cite{lamoreaux_solid-state_2002}. Applying the magnetic field could lead to the EDM-induced electrization which opens an opportunity to measure it.  By reciprocity, applying  instead a strong electric field  to a sample of paramagnetic atoms could lead to the sample magnetization under right conditions.

An atom with an electric moment $d$ has the energy $-\vec{d}\cdot\vec{E}$ in the external electric field $\vec{E}$. A sample of such atoms acquires an induced magnetic field $\vec{B}_E$, since the alignment of atomic spins minimizes this extra energy; clearly, the thermal motion reduces their ability to maintain such an alignment. The generated magnetic flux through the surface area $A$ and corresponding field are estimated as \cite{lamoreaux_solid-state_2002}
\begin{equation}
	\Delta \Phi = 4\pi \chi A \frac{d}{\mu} E\qquad\to\qquad B^{I}_e = \frac{\Delta \Phi}{A} = 4\pi \chi  \frac{d}{\mu} E\,,
\end{equation}
where the magnetic susceptibility $\chi= \rho \mu^2/(3k_B T)$ is temperature-dependent, $\rho$ is the atom density, the superscript means the first contribution into the induced magnetic field. If atoms have cubic arrangement, the susceptibility is mainly determined by its scalar part \cite{baryshevsky_time-reversal-violating_2004}.  

It is the primary mechanism of magnetic field generation by external electric field which is enabled by the existence of atomic EDM $d\ne0$. The additional contribution is enabled by existence of $P$- and $T$-odd polarizabilities that we discussed before, see the right-hand sides in \eqref{eq6}. The second contribution is then given as 
\begin{equation}\label{eq16}
	(B_e^{II})_i  = 4\pi \rho \alpha^\prime_{ik}   E_k\,.
\end{equation}
Technically, this $P$- and $T$-odd polarizability is determined by mixed atomic matrix elements, see their estimates in \cite{baryshevsky_time-reversal-violating_2004} and references therein; it arises due to P- and T-odd interaction of electrons with the nucleus. If shifts of atomic levels caused by external fields are comparable with the interaction energy, then the polarizability $\alpha^\prime_{ik}(E)$ becomes field-dependent.  The overall magnitude of induced magnetic field is the sum of these two contributions
\begin{equation}\label{eq13}
	\vec{B}_e = \vec{B}_e^{I}+\vec{B}_e^{II}+\dots  = 4\pi\frac{\chi}{\mu} \qty(d + \frac{\rho\mu}{\chi} \alpha^\prime+\dots)   \vec{E}\,,
\end{equation}
where by dots we show some other contributions that might be important depending on sample geometry and other factors, see details in \cite{baryshevsky_time-reversal-violating_2004}. We can immediately comment that this effect will exist even for zero EDM coefficient if the primed polarizability is nonzero.

By reciprocity, if an external magnetic field is applied to such a sample, initially disoriented spins are forced to align which in turn aligns their electric dipole moments. Therefore, the electric moments that now become aligned at macroscopic scale induce the electric field $\vec{E}_B$\hspace{-3pt}
	\footnote[7]{D.DeMille in \cite{lamoreaux_solid-state_2002}}
\begin{equation}\label{eq7}
	E_b^I = 4\pi \rho d P(B) \,,
\end{equation}
where the function $P(B)$ takes into account both degree of spin polarization and particular sample geometry \cite{baryshevsky_time-reversal-violating_2004}.  It is again the primary mechanism of electric field generation by external magnetic field which is enabled by the existence of atomic EDM $d\ne0$. Since there exists the second type of $P$- and $T$-odd polarizability $\beta^\prime_{ik}$, see \eqref{eq6}, the additional contribution is given by 
\begin{equation}\label{eq8}
	(E_b^{II})_i  = 4\pi \rho \beta^\prime_{ik}   B_k\,.
\end{equation}
Similarly to $\alpha^\prime_{ik}$, this tensor polarizability is determined by mixed matrix elements. Effectively, since $\alpha^\prime_{ik}$ and $\beta^\prime_{ik}$ are interrelated \cite{baryshevsky_time-reversal-violating_2004, baryshevsky_high-energy_2012}, we are dealing with one $P/T$-odd tensor quantity in  \eqref{eq16} and \eqref{eq8}. Combining both contributions, the magnitude of induced electric field is given by
\begin{equation}\label{eq14}
	\vec{E}_b = \vec{E}_e^I+\vec{E}_b^{II} = 4\pi \rho\qty[d\, P(\vec{B}) +\beta^\prime \vec{B}] \,,
\end{equation}
where we see that the similar structure emerges again. 

Measurements of EDM coefficients by detecting the induced electric or magnetic field require the estimates of the primed polarizabilities in addition to other pertinent factors. Since both $\chi$ and $P(B)$ are temperature-dependent while $(\alpha^\prime_{ik}, \beta^\prime_{ik})$ are not, measuring the induced fields at different temperatures allows to evaluate the EDM contribution in principle.  An analysis of experimental limitations of measuring magnetic fields that are induced by such a mechanism is given in \cite{budker_sensitivity_2006}, while the experimental results are reported in \cite{eckel_limit_2012}.

\section{Spin one particles}
Next, let us consider the case of spin one particle moving in an electromagnetic field $(\vec{E}, \vec{B})$. Our goal here is to trace out the derivation of the spin motion to see how the EDM related terms and the other terms that accompany it emerge. Hence, we include the relevant terms only, and omit other terms that are not relevant for the current discussion. The detailed derivation is given in \cite{baryshevsky_rotation_2008} and references therein. 

We consider now the polarization of deuteron beam for the conditions of storage ring experiments. The existence of deuteron EDM can lead to a growth in beam vertical polarization for configurations where its magnetic moment causes the spin precession only in the horizontal plane \cite{baryshevsky_phenomenon_2004}. One would consider it as an ideal set-up where the EDM  influence is not overshadowed  by other factors. However, as we are going to see again, and it follows the scope of this review, that other effects  imitate the influence of deuteron EDM. Here, these other factors are deuteron electric and magnetic polarizabilities which unavoidable contributions into the dynamics of vertical polarization must be accurately taken into account \cite{baryshevsky_birefringence_2005, baryshevsky_spin_2005, baryshevsky_rotation_2008}. 

The starting point is the quantum-mechanical equation for the wave function $\psi(t)$
\begin{equation}\label{eq1}
	i\hbar\frac{\partial\psi(t)}{\partial
	t}=\left(\hat{H}_{0}+\hat{V}_{edm}+\hat{V}_{\vec{E}}+\hat{V}_{\vec{B}}+\dots\right)\psi(t) \,,
\end{equation}
where $H_0$ is the part of total Hamiltonian that leads to the conventional Bargmann-Mitchel-Telegdi (BMT) equation, and $\psi(t)$ belongs to  multiplet $S=1$. The dots represent terms that are specific to a particular experimental set-up: for example, they might describe interactions with target nucleons or other fields; such additional terms are not shown here.  The second term in \eqref{eq1} describes the particle EDM interaction with external field
\begin{equation}
	\hat{V}_{edm} =
	-\vec{d}\cdot \left(\vec{E}+\vec{\beta}\times\vec{B}\right) = - \vec{d} \cdot \vec{E}^\prime,
\label{edm}
\end{equation}
where the EDM operator is $\vec{d} = d \vec{S}$. Three spin operators $S_i$ are defined by means of $3\times 3$ matrices of $SO(3)$ representation. Two vectors of effective fields are defined as 
\begin{gather}
\begin{aligned}
&\vec{E}^\prime &&= \vec{E}+\vec{\beta}\times\vec{B} &&= \abs{E^\prime}\, \vec{n}_E\,,\\[1ex]
&\vec{B}^\prime &&= \vec{B}-\vec{\beta}\times\vec{E} &&= \abs{B^\prime}\, \vec{n}_B\,,
\end{aligned}
\end{gather}
hence two unit vectors, $\vec{n}_E$ and $\vec{n}_B$, define the directions of effective fields respectively. They are the fields in the particle rest frame which rotates together with the particle relative to the laboratory frame. 

Two remaining terms in the right-hand side of \eqref{eq1} correspond to polarization energy that is induced  by external field
\begin{gather}
\begin{aligned}
	&\hat{V}_{E}&&=-\frac{1}{2}\hat{\alpha}_{ik}E^\prime_{i}E^\prime_{k}&&=
	\alpha_{s}E^{\prime 2}-\alpha_{t}E^{\prime2}\left(\vec{S}\vec{n}_{E}\right)^{2}\,,\\[1ex]
	& \hat{V}_{B}&&=-\frac{1}{2}\hat{\beta}_{ik}B^\prime_{i}B^\prime_{k}&&=\beta_{s}B^{\prime 2}-\beta_{t}B^{\prime 2}\left(\vec{S}\vec{n}_{B}\right)^{2},
\end{aligned}
\end{gather}
where $\hat{\alpha}_{ik}$ and $\hat{\beta}_{ik}$ are operators of magnetic and electric polarizabilities respectively; $\alpha_s$ and $\beta_s$ are their scalar parts, while $\alpha_t$ and $\beta_t$ are the tensor ones. 

The next step is to obtain the equations for observables that follow from  \eqref{eq1}. Following the approach of \cite{baryshevsky_rotation_2008} and references therein, the motion equation of deuteron spin $\vec{s}$ is obtained from \eqref{eq1} as
\begin{equation}\label{eq2}
\frac{d\vec{s}}{dt}=
\vec{s}\times \vec{\Omega}
 +
d(\vec{s}\times \vec{E}^\prime)-\frac{2}{3}\alpha_{T}E^{\prime 2}[\vec{n}_{E}\times\vec{n}_{E}^{\prime}]
-\frac{2}{3}\beta_{T}B^{\prime 2}[\vec{n}_{B}\times\vec{n}_{B}^{\prime}],\\
{} 
\end{equation}
where the Thomas-BMT precession $\vec{\Omega}$ is given by the regular expression, see equation (4) from \cite{baryshevsky_rotation_2008}, for example. It includes the anomalous magnetic moment too. The last two terms that complicate the spin dynamics are expressed by means of the polarization tensor $P_{ik}$
\begin{align}
	&\vec{n}_{E}^{\prime} = P \vec{n}_{E}\,,		&\vec{n}_{B}^{\prime} = P \vec{n}_{B}\,,
\end{align}
where the traceless and symmetric $3 \times 3$ tensor $P_{ik}$ is quadratic in spin matrices $S_i$. Its evolution equation is given as
\begin{multline}\label{eq3}
	\frac{dP_{ik}}{dt}  =
	-\left(\varepsilon_{jkr}P_{ij}\Omega_{r}+\varepsilon_{jir}P_{kj}\Omega_{r}\right)-
	\frac{3}{2}\alpha_{t}E^{\prime 2}\{(\vec{n}_{E}\times\vec{s})_{i}, (\vec{n}_{E})_k\}\\[1ex]
	-\frac{3}{2}\beta_{t}B^{\prime 2}\{(\vec{n}_{B}\times\vec{s})_{i}, (\vec{n}_{B})_k\}\,,
\end{multline}
here $\Omega_r$ are components of vector $\vec{\Omega}$.

The equations \eqref{eq2} and \eqref{eq3} form the system of linear equations for components of both spin vector $\vec{s}$ and polarization tensor $P_{ik}$. Remarkably, the solution can be given for few selected arrangements. In the coordinate system that rotates together with the particle velocity $\vec{v}$, the spin will rotate with frequency $\frac{e a}{m} B$ determined by the particle anomalous magnetic momentum $a$.\hspace{-4pt}
	\footnote[7]{This idealized picture is valid if the magnetic polarizability is ignored, see below. In such a case, the orbits are circular.}
In such a coordinate system, the particle velocity $\vec{v}$ remains orthogonal to both $(\vec{E}, \vec{B})$; we can conveniently select $\vec{B}$ aligned along the axis $z$, while $\vec{E}$ is directed along $y$. This arrangement locks the motion induced by deuteron magnetic moment in the $(x,y)$ plane. Then the EDM should lead to spin motion in the $z$ direction which opens the opportunity of measuring the EDM coefficient. 
Ignoring first the magnetic polarizability, the solution of \eqref{eq2} and \eqref{eq3} is given as
\begin{equation}\label{eq4}
	\dv{s_3}{t} = -\qty( d s_2+\frac{2}{3}\alpha_{t}E^\prime\, P_{12}  ) E^\prime\,,
\end{equation} 
which shows that the vertical beam polarization is influenced by both deuteron EDM and electric polarizability component $\alpha_t P_{12}$. Taking into account the magnetic polarizability now, the solution shows that the vertical spin component receives an additional contribution  $\dot s_3 {\sim} \sin^2(\frac{e a}{m} B t)$. The horizontal spin will also rotate with two frequencies around $\vec{B}$ instead of the single one as we mentioned before for the idealized situation. In this case, the symmetry-violating polarizabilities have not been included into the analysis; even the regular ones are shown to interfere with discerning the contribution of intrinsic EDM from other factors that mimic  its influence on particle spin. Inserting a hypothetical $P/T$-odd polarizability of deuteron into the analysis will add extra terms into the right-hand side of \eqref{eq4}.

Therefore, both deuteron EDM and its polarizability contribute into the spin dynamics under the reviewed storage ring conditions. It does not preclude unambiguous EDM measurements however. The contributions from both magnetic and electric polarizabilities depend on external field, and allow for unambiguous identification, provided the experiments are sufficiently sensitive and take the additional factors into account. These options are thoroughly discussed in \cite{baryshevsky_rotation_2008}; additional considerations for amplifying the EDM contribution relative to the polarizability effect are also given in  \cite{silenko_tensor_2007}.

\section{Electron EDM in storage rings}
The cases that we have reviewed so far involve atoms and deuteron which are composite particles. It could have been argued that in dealing with elementary particles which are not trapped in bound states  the interplay between nonzero EDM and $P/T$-odd polarizabilities would be avoided. The key here, however, is the mutual existence of intrinsic EDM itself and some other pseudoscalar quantity that is able to mix electric and magnetic contributions. An appearance of such quantities in equations for observables is the direct consequence of allowing for symmetry-violating interactions.  Remember that the scalar parts of $P$ and $T$-odd polarizabilities $(\alpha^\prime_{ik}, \beta^\prime_{ik})$ must be pseudoscalars to compensate for the pseudoscalar nature of product $\vec{E}\cdot\vec{B}$, see \eqref{eq9}. Remarkably, such a pseudoscalar emerges naturally in deriving the BMT-like equation from the Dirac one if the symmetry-violating EDM term is included. 

The BMT equation and related equations in different frames can be obtained in a variety of ways, see for example \cite{berestetskii_quantum_1982, silenko_quantum-mechanical_2005,fukuyama_derivation_2013}.  However there exists only  one derivation \cite{rafanelli_classical_1964}  known to us that explicitly deals with the pseudoscalar density $i\bar{\psi}\gamma^5\psi$.   It is based on the WKB approach to Dirac equation.  To the best of our knowledge, in all other existing derivations, the pseudoscalar is not even mentioned; hence it has been effectively set to zero in such derivations. Having in mind the search for $CP$-violating interactions where pseudoscalar currents might play a prominent role, we do not make this assumption here. 

Following \cite{rafanelli_classical_1964} and \cite{khriplovich_cp_1997}, we add two terms to the Dirac Lagrangian 
\begin{equation}
	\mathcal{L} = \psi(i \slashed D - m )\psi -\frac{a\,e}{4m}F_{\mu\nu}\bar\psi\sigma^{\mu\nu}\psi-i \frac{d }{2}F_{\mu\nu}\bar\psi\sigma^{\mu\nu}\gamma^5\psi\,,
\end{equation}
where $\slashed D$ is the gauge covariant derivative, $a$ and $d$ are the anomalous magnetic  and electric dipole moments respectively, and $F_{\mu\nu}$ is the tensor of electromagnetic field. Correspondingly, the equation for Dirac wave functions follows as
\begin{equation}\label{chFields:eq1967}
	\qty[i \slashed \partial -  e \slashed A_\mu-m - (\frac{a e}{2m}+id \gamma^5)\frac{\sigma^{\mu\nu}}{2}F_{\mu\nu}] \psi = 0\,. 
\end{equation}
In \cite{porshnev_electron_2020}, we followed the  WKB-based derivation \cite{rafanelli_classical_1964} which was straightforwardly extended to include the EDM term. The critical point in derivation comes at the very end where \cite{rafanelli_classical_1964} explicitly sets the Dirac pseudoscalar bilinear $q=i\bar\psi \gamma^5\psi$ to zero. We did not make this assumption following our expectation that such a quantity might play an important role for in the context of this study, very similar to the $P$ and $T$-odd polarizabilities used before. The spin precession equation was obtained as
\begin{multline}\label{eq11}
	\dv{s^\rho}{\tau} 	=\qty(\frac{ge}{2m}+2d\frac{q}{r}  ) F^{\rho\nu}s_\nu
	+\qty[\frac{a\,e}{m}(1-\frac{q^2}{r^2})+2d\frac{q}{r} ] s^\mu F_{\mu\nu}u^\nu u^\rho\\[1ex]
	- \qty( 2 d- \frac{a\,e}{m} \frac{q}{r}) \tilde F^{\rho\nu}  s_\nu
	- \qty[2 d(1 -\frac{q^2}{r^2}) -\frac{a\, e }{m}\frac{q}{r} ] s^\mu \tilde F_{\mu\nu}  u^\nu u^\rho  \,,
\end{multline}
where $g=2$ is magnetic $g$-factor, $\tau$ is proper time, $s_\mu$ is the spin four-vector, $u_\mu$ is electron four-velocity. The scalar density $r=\bar\psi \psi\approx 1$ is expected to be close to unity.  The pseudoscalar $q$ intermixes magnetic and electric moments in \eqref{eq11}, similarly to the $P$ and $T$-odd polarizabilities in the non-relativistic cases reviewed before.  Setting the pseudoscalar $q$ to zero turns \eqref{eq11} exactly  into the well known BMT equation with EDM term included \cite{fukuyama_derivation_2013}.

Since the estimate \cite{porshnev_electron_2020} shows that the pseudoscalar-induced corrections to magnetic terms are too small, at least in the electron case, the equation \eqref{eq11} can be simplified as
\begin{equation}\label{eq12}
	\dv{s^\mu}{\tau} 
		=\frac{ge}{2m} F^{\mu\nu}s_\nu+\frac{a\, e}{m}( s^\rho F_{\rho\nu}u^\nu)  u^\mu	- d^{\prime}\qty( \tilde F^{\mu\nu}  s_\nu+  s^\rho \tilde F_{\rho\nu}  u^\nu u^\mu )  \,,
\end{equation}
where the effective EDM is 
\begin{equation}\label{eedm_10}
	d^{\prime} = 2 d -\frac{q}{r} \frac{a\, e }{m} \,.
\end{equation}
Remarkably, the WKB-based derivation yields the same functional form of the BMT equation that was derived in \cite{fukuyama_derivation_2013}, and many other sources.  Therefore, we can directly apply its other forms, derived in many works of Fukuyama, Silenko and others for comparison with experimental data in laboratory frames, including the rest frame.  These forms properly describe the Thomas precession and other relativistic effects in relevant frames. To take into account the pseudoscalar correction, their Lorentz scalar $2d$ must be replaced with our Lorentz scalar $2d - q a_e e/m$ in any of these other forms.

An immediate question can be raised why would the pseudoscalar $i\bar\psi \gamma^5\psi$ differ from zero for free or quasi-free electrons.  Remember that nonzero $P$ and $T$-odd polarizabilities which also have pseudoscalar component are explained by mixing of atomic orbitals with different parity. What would be a realistic physical model in the free or quasi-free electron case?  Consider, for example,  a linear superposition of particle  and antiparticle 
\begin{equation}
	\psi = A \mqty(\chi\\ 0) + B \mqty(0\\ \xi)\,,
\end{equation}
where we use the standard representation of Dirac gammas; $\chi$ and $\xi$ are undotted and dotted unit spinors respectively. The pseudoscalar estimates to
\begin{equation}
	q = i\bar\psi \gamma^5\psi = i (A^* B \chi^\dagger\xi - AB^* \xi^\dagger\chi )\,.
\end{equation}
If the Dirac fermion is in a pure particle $(B=0)$ or pure antiparticle state $(A=0)$, then $q$  is strictly zero. For the  fermion to possess a nonzero pseudoscalar, it must have some admixture of antifermion component. This assumption is supported by our understanding of realistic electrons. First, the ideal Dirac electron does not even possess the anomalous magnetic moment which we had to insert into the Dirac equation by hand. Second, the vacuum polarization around any bare charge creates virtual pairs which will add the required component to  bare Dirac electrons. 

Let us pause for a second to elaborate on the role of pseudoscalar, which is one of sixteen gauge-invariant Dirac bilinears.  The complete system of equations for these bilinears that follow from the Dirac equation is complicated, and cannot be given in the closed form; see however \cite{inglis_fierz_2014}. Only individual equations, like the divergence of axial current \cite[p. 51]{peskin_introduction_1995}, can be given in a closed form.  The WKB allows the simplified system to be obtained in the closed form
\begin{equation}
	\begin{rcases}
		\partial_\mu u^\mu  &=\phantom{-}0\,,\\
		\partial_\mu s^\mu  &=\phantom{-}2 m q\,,\\
		\partial^\mu q		&=-2 m s^\mu-\dots\,,\\
		\dots
	\end{rcases}	\qquad\overset{\text{WKB}}{\to}\qquad
	\begin{cases}
			\dot{r}	&=0\,,\\
			\dot{q}	&=0 \,,\\
			\dot{u}^\mu &=\frac{e}{m}F^{\mu\nu} u_\nu\,,\\
			\dot{s}^\mu &=\text{BMT terms with  }d^\prime\,,
	\end{cases}
\end{equation}
It includes only ten Dirac bilinears however: density $r=\bar{\psi}\psi$, pseudodensity $q = i\bar\psi \gamma^5\psi$, current $u^\mu{\sim}\bar\psi \gamma^\mu\psi$ and spin $s^\mu{\sim}\bar\psi \gamma^\mu\gamma^5\psi$.  The first two are the constants of motion with values that must be set outside the first order WKB approximation. This is the key point: setting the pseudoscalar $q$  to a certain value is an assumption that is independent of the WKB derivation of BMT equation. 

The first WKB order simply says that two densities are constants of motion that are not influenced by four-velocity, spin or weak EM fields in this approximation; similarly, the four-velocity $u^\mu$ does not depend on $s^\mu$ in the lowest order. However, the WKB does not say what value q has.  Hence, there are two options:
\begin{itemize}
\item set $r=1$ and $q=0$ (it leads to the classical BMT case)
\item allow for q be nonzero (it is not forbidden by any other laws)
\end{itemize}
For the second option, we have to look outside the WKB to figure out a value of q.  Since $q$ is expected to be much smaller than the regular density $r$, it is really a minute and non-essential point for most practical applications. However, the current upper bound on  electron EDM is extremely low \cite{andreev_improved_2018}:  if $q$ is anywhere close to $10^{-15}$ (comparing with density that is one), it will be a dominant correction into the effective electron EDM $d^\prime$ according to the WKB solution. Taking into account that many extensions to the standard model \cite{pospelov_ckm_2014,cesarotti_interpreting_2019,flambaum_sensitivity_2020} consider terms with nonzero pseudoscalar currents, it will be great to continue investigating the potential impact of nonzero $q$ on spin dynamics.

Comparing now the effective EDM $d^\prime$ given by \eqref{eedm_10} with the effective EDM $d^\prime$ for the experiments with atomic systems, see \eqref{eq13} and \eqref{eq14}, we see the remarkable analogy. In both cases, the intrinsic EDM coefficient receives the pseudoscalar correction that is proportional to the magnetic moment. More, in both cases this correction is of dynamic nature and depends on external factors. Concluding, the result from \cite{porshnev_electron_2020} is just another example of the feature of EDM-induced dynamics that was discovered and described in \cite{baryshevsky_phenomenon_1999, baryshevsky_time-reversal-violating_2004} many years ago.

\section{Conclusions}
We have reviewed several different experiential set-ups for measuring the EDM coefficient. Our review shows that the EDM-induced dynamics is unavoidably controlled by the intrinsic EDM itself and other $P$- and $T$-odd quantities, in addition to other factors that are particular to a given experimental set-up. It makes possible a situation when even if the EDM constant $d$ is zero, nonzero pseudoscalar quantities will make the system behave in electric field as it possesses an electric dipole moment.  For atomic systems, these additional factors are originated by mixing of atomic orbitals with different parities.  Examples of such a mixing that influences various $P$- and $P/T$-violating phenomena are given in \cite{khriplovich_parity_1991, khriplovich_cp_1997}. For particles, it is studied in QFT models that extend the standard model by introducing $CP$-violating terms \cite{cesarotti_interpreting_2019,altmannshofer_electric_2020}. The terms they typically consider include the product of pseudoscalar  currents with some other currents. Hence, it is expected that such extensions of standard model will lead to both nonzero EDM and pseudoscalar quantities. 

The corrections to intrinsic EDM coefficients are of dynamic nature, since they depend on external factors. In principle, it makes the contributions of intrinsic moments distinguishable from the effects of field-dependent polarizabilities and other dynamic factors. The regular experimental tricks to suppress these factors are to change the relative field orientation, average data in space and time, or adjust field frequencies. This review allows to suggest an additional way to improve the sensitivity of EDM experiments  or rather strengthen arguments in selecting proper objects for such experiments.

First, the correction to EDM has the structure of pseudoscalar times magnetic moment. It is explicitly shown in \eqref{eedm_10}, and it is also known for the  $P$ and $T$-odd polarizabilities $(\alpha^\prime_{ik}, \beta^\prime_{ik})$, see \cite{baryshevsky_parity_1993}. Since the magnetic moment is inversely proportional to particle mass, by selecting the heaviest particle, the correction is expected to scale down relative to the intrinsic EDM. Second, the mass scaling for particle EDMs is still an open question.  Many scalings \cite{grozin_upper_2009,engel_electric_2013, panico_eft_2019}  show that the intrinsic EDM scales linearly with particle mass. For examples, three crude estimates of electron EDM are given as \cite{panico_eft_2019}
\begin{equation}
	\frac{d_e}{e} \simeq \qty(\frac{g^2}{16\pi^2})^2  \frac{m_e}{\Lambda^2}\,,\qquad\quad \frac{d_e}{e} \simeq \qty(\frac{g^2}{16\pi^2})^2  \frac{m_e m_W^2}{\Lambda^4}\,\qquad\quad \frac{d_e}{e} \simeq \frac{1}{8\pi^2}  \frac{m_e}{f^2}
\end{equation}
where $g$ is the electroweak coupling, $m_W$ is the $W$-boson mass, $\Lambda$ is the cutoff energy which might define the scale of new physics, and $f$ is the Higgs decay constant. The similar scaling for proton can be found in  \cite{semertzidis_storage_2016}. Many refined estimates of electron EDMs from \cite{cesarotti_interpreting_2019, panico_eft_2019} retain this dependence on particle mass, while there also exist estimates of intrinsic $d_e$ that do not appear to be explicitly dependent on electron mass. 
\begin{table}
\centering
  \begin{threeparttable}
\caption{Upper bounds on electric dipole moments (C.L. 90\%) and their ratios to anomalous magnetic moments for fermions.}    
\begin{tabular}{lccccccccccccccc}
\hline
\hline
Particle	& m  	&  $\abs{d_e/e}$			& $\abs{a_e}$	&&Ratio\\
			& $(MeV)$	&	$cm$			& 	&&$m d_e/a_e$	\\
\hline
electron	&$0.51$	&$1.1\cross10^{-29}$	& $1.2\cross 10^{-3}$	&& $4.7\cross 10^{-27}$		\\
muon     	&$105.7$		&$1.5\cross 10^{-19}$	&$1.2\cross 10^{-3}$   && $1.3\cross 10^{-14}$ 		\\
tau			&$1777.$		&$1.6\cross 10^{-18}$ 	&$1.1\cross 10^{-3}$	&& $2.6\cross 10^{-12}$		\\[1ex]
neutron		&$940.$		&$1.8\cross 10^{-26}$ &	$0.044$	&& $3.8\cross 10^{-22}$	\\
proton		&$939.$		&$1.7\cross 10^{-25}$ &	$0.397$	&&$4.0\cross 10^{-22}$\\
$\Lambda^+$	&$1115.$	&$1.5\cross 10^{-16}$ &	$0.693$	&&$2.4\cross 10^{-13}$	\\
\hline

\end{tabular}
\label{eedm_table}
\begin{tablenotes}
      \footnotesize
      \item Data from \cite{andreev_improved_2018, muon_g-2_collaboration_improved_2009, inami_search_2003,blinov_upper_2009, zyla_review_2020}. Weak $a_e$ for tau.
    \end{tablenotes}
  \end{threeparttable}
\end{table}

Therefore, the ratio of intrinsic EDM to the $P/T$-odd correction that mimic EDM is expected to scale with particle mass as
\begin{equation}
	\frac{d_e}{\langle P/T\text{-odd correction}\rangle}\sim\frac{m d_e}{a_e}\sim m^\alpha\,,
\end{equation}
where $\alpha\sim2$, and $a_e=(g-2)/2$. Remarkably, this ratio for electron was already discussed \cite{commins_electric_2009} in the different context
\begin{equation}
	\frac{d_e}{(g-2)\mu_B}\sim \qty(\frac{m_e}{m_x})^2\,,
\end{equation}
where $m_x$ is a hypothetical heavy particle which interaction with electrons leads to a $CP$-violating phase. It was proposed there as a crude universal estimate for electron EDM that should be expected in extensions of the standard model. We view this ratio instead as a crude measure of relative strengths of intrinsic or static  EDM $d_e$ to the dynamic EDM correction that is also originated by $CP$-violating interactions. Our analysis and the obtained expression \eqref{eedm_10} allows us to hypothesize that this scaling should be expected for other fermions which are describable by Dirac equation in the classical limit. The corresponding estimates are summarized in table \ref{eedm_table}. The ratio is significantly higher for heavier particles, however it also reflects the different levels of uncertainty in the upper bounds of corresponding $d_e$ for unstable or difficult-to-handle fermions. Anyway, selecting the heaviest particles for EDM experiments,  e.g. $\Lambda$-baryons or similar ones, is expected to  provide additional benefits in isolating the contribution of intrinsic $d_e$ from other contributions that are originated by $P/T$-odd interactions. Clearly, it is just one factor is deciding which experimental set-up is more favorable for a first detection of $d_e$, since other experimental factors might be more significant for the overall experimental sensitivity. From the point of this analysis, the experiments with heavier baryons in bent crystals \cite{baryshevsky_spin_2015, baryshevsky_search_2017, baryshevsky_electromagnetic_2019, baryshevsky_electromagnetic_2019-1,botella_search_2017, bagli_electromagnetic_2017-1 }
have the additional edge in measuring nonzero EDM that has been escaping the detection so far.

\ack
One of the authors (PP) thanks M.~Pospelov for the stimulating discussion regarding the Dirac pseudoscalar. 

\section*{References}
\bibliographystyle{unsrt}

\end{document}